\documentclass[9pt,twocolumn,twoside]{osajnl}

\journal{josab} 

\setboolean{shortarticle}{false} 

\ifthenelse{\boolean{shortarticle}}{\colorlet{color2}{color2b}}{\colorlet{color2}{color2}} 

\title{Revisiting the boundary conditions for second-harmonic generation at metal-dielectric interfaces}

\author[1,*]{K. Nireekshan Reddy}
\author[1,2]{Parry Y. Chen}
\author[3]{Antonio I. Fern\'andez-Dom\'inguez}
\author[1]{Yonatan Sivan}

\affil[1]{Unit of Electro-Optic Engineering, Ben-Gurion University, Beer-Sheva, 8410501, Israel}
\affil[2]{School of Physics and Astronomy, Raymond and Beverly Sackler Faculty of Exact Sciences, Tel Aviv University, Tel Aviv, 69978, Israel}
\affil[3]{Departamento de F\'isica Te\'orica de la Materia Condensada and Condensed Matter Physics Center (IFIMAC), Universidad Aut\'onoma de Madrid
E-28049 Madrid, Spain}

\affil[*]{Corresponding author: kothakap@post.bgu.ac.il}

\dates{Compiled \today}

\ociscodes{(190.0190) Nonlinear optics; (190.4350) Nonlinear optics at surface; (240.6680) Surface plasmons; (030.4070) Modes; (190.0190) Resonance; (050.1755) Computational electromagnetic methods.}

\doi{\url{http://dx.doi.org/10.1364/ao.XX.XXXXXX}}

\begin{abstract}
We study second-harmonic generation (SHG) arising from surface nonlinearity at a metal-dielectric interface using a spectral decomposition method. Since 
our method avoids the need to consider the generalized boundary condition across the metal-dielectric interface 
in the presence of a perpendicular surface source, we 
retrieve the 
known discontinuity of the tangential component of the electric field ($\mathbf{E}_{\parallel}^{2\omega}$) for a general geometry, 
based on a purely mathematical argument. Further, we reaffirm the standard convention of the implementation of this condition, namely, that the surface dipole source radiates as if placed outside the metal surface for arbitrary geometries. We also study and explain the spectral dependence of the discontinuity of the tangential component of the electric field at second harmonic. Finally, we note that the default settings of the commercial numerical package COMSOL Multiphysics fail to account for the $\mathbf{E}_{\parallel}^{2\omega}$-discontinuity. We provide a simple recipe that corrects the boundary condition within these existing settings. 
\end{abstract}

\setboolean{displaycopyright}{true}

\begin{document}

\maketitle
\thispagestyle{fancy}

\ifthenelse{\boolean{shortarticle}}{\ifthenelse{\boolean{singlecolumn}}{\abscontentformatted}{\abscontent}}{}

\section{Introduction}\label{sec:sec1}
Second-order optical nonlinear processes in centrosymmetric materials have been studied since the early days of nonlinear optics~\cite{blm1,franken,pershan,blm2,adler,jha,sgh_first_exp}. It is well known that these processes are symmetry forbidden in the local bulk response limit, and thus are governed by higher-order nonlocal bulk and surface symmetry breaking effects~\cite{blmrev,rudnik,sipe80}. Recently, there has been renewed interest in surface nonlinearities from metal-dielectric interfaces, in which electric fields can be greatly enhanced due to plasmon resonances. Specifically, advances in fabrication of nanostructures have allowed the experimental investigations of second-harmonic generation (SHG) from single metallic nanoparticles~\cite{sn1,sn2,sn3,sn4,sn5,jnl}, split-ring resonators~\cite{srr1,srr2}, periodic nanostructured metal films ~\cite{prdst1}, and other  nano-geometries~\cite[see Ref.][and references therein for a more thorough review]{zayat_rev}. The corresponding theoretical modeling relied either on surface nonlinearity~\cite{butet_coreshell,bem,martan} or non-local bulk nonlinearity~\cite{pavel1,pavel2}, while some studies have incorporated both effects~\cite{dadap1,mendez1,mendez2,dadap2,sdmshg1,sdmshg2,scaloraPRA,ciraci1,ciraci2,miano}. Importantly, even the bulk nonlinearity can be mapped to a surface current source, so that the surface nonlinearity provides a simple, general model for second-order nonlinear phenomena in metals~\cite{sipe_srf_map,miano,ciraci1,ciraci2}. 

In Ref.~\cite{hnz}, Heinz showed that a {\em non-uniform, normal} component of the surface current 
results in a discontinuous tangential electric field ($\mathbf{E}_{\parallel}$) across the interface between the two media. Computing the nonlinear polarization at the second-harmonic (SH) can involve the normal component of the electric field at the fundamental frequency (FF). Since the normal component of the electric field is discontinuous across the interface, it was \textit{a-priori} not clear whether one has to choose the fields inside or outside the metal for evaluating the SH electromagnetic fields accurately. This was resolved by Sipe \textit{et al.} as follows: one has to evaluate the nonlinear polarization based on the fundamental fields {\em inside} the metal; the dipole layer source generated from the nonlinear polarization should then be placed just {\em outside} the metal surface~\cite{sipe80}.

Here, we revisit the derivation of the generalized boundary condition (BC) in the presence of surface sources and provide an alternative description of surface SHG using the spectral decomposition method (SDM) in a full electrodynamic setting~\cite{DJB}. We show that the generalized BC is naturally reproduced in the SDM formulation without invoking Heinz's generalized BC~\cite{hnz}. Based on SDM, we show from a purely mathematical point of view and for a general geometry that the dipole layer radiates indeed as if placed outside the metal surface, in agreement with Sipe \textit{et al.}~\cite{sipe80}. As a specific example, we apply SDM to an infinite metallic cylinder under normal incidence and provide physical insight into the spectral dependence of the discontinuity of the parallel component of the SH electric field, $\textbf{E}_{\parallel}^{2\omega}$. In particular, we show that the $\textbf{E}_{\parallel}^{2\omega}$-discontinuity is maximal away from the SH resonance, yielding 50\%--100\% errors if not implemented correctly; yet, the discontinuity is smaller at a SH resonance, its value diminishing as the quality factor of the resonance improves. Finally, we note that the default settings of COMSOL Multiphysics, a commonly used commercial solver of Maxwell's Equations, neglect the normal component of the surface current $J_{S,\perp}^{(2)}$ and fail to account for the $\mathbf{E}_{\parallel}^{2\omega}$-discontinuity. As a remedy, we provide a simple recipe that accounts for the error within the existing settings using the insight provided by SDM.

The paper is organized as follows: In  Sec.~\ref{sec:sec2} we review the generalized BCs for electromagnetic waves and their implementation. Sec.~\ref{sec:sec3}  briefly introduces the  spectral decomposition method (SDM) and its application to SHG.  We calculate the SHG from a metallic infinite cylinder with a surface nonlinearity using SDM in Sec.~\ref{sec:sec4} and study the spectral dependence of the $\textbf{E}_{\parallel}^{2\omega}$-discontinuity  field in Sec.~\ref{sec:sec5}. An alternative approach for the numerical implementation (in COMSOL Multiphysics) of the generalized BC is discussed in Sec.~\ref{sec:sec6}. We conclude with a discussion and outlook.

\section{GENERALIZED BOUNDARY CONDITIONS AND THEIR IMPLEMENTATION}\label{sec:sec2}
Consider a surface SH polarization vector at a metal-dielectric  interface given by
\begin{eqnarray} \label{p2_gen}
\mathbf{P}^{(2)}(\mathbf{r}) = \mathbf{P}^{(2)}_S({\mathbf{r}}_{\parallel}) \delta(r_{\perp}),
\end{eqnarray}
where $\delta({{r}}_{\perp})$ is the Dirac-delta function~\cite{hnz}. Note that we denote the vector and scalar quantities in bold and italic font, respectively. The subscript $\parallel$ ($\perp$) corresponds to the direction tangential (normal) to the interface. For example, for a flat interface at $z=0$, one would have $\mathbf{r}_\parallel = (x,y)$ and $r_\perp = z = 0$. The SH polarization~(\ref{p2_gen}) can also be formulated as a surface current density, $\mathbf{J}_S^{(2)} = - 2 i \omega \mathbf{P}_S^{(2)}$, or as surface charge  $\varsigma^{(2)}$ or volume charge $\varrho^{(2)}$ densities, see Appendix~\ref{apenA_sources}. 

In the standard formulation of the Maxwell´s Equations for SH 
fields, $\mathbf{P}^{(2)}_S$ appears only in the BCs at the metal-dielectric interface as an external source. 
The tangential component of $\mathbf{P}_S^{(2)}$, denoted by $\mathbf{P}_{S,\parallel}^{(2)}$, gives rise to a surface current $\mathbf{J}_{S,\parallel}^{(2)}$ tangential to the interface. Surface currents $\mathbf{J}_{S,\parallel}^{(2)}$ are very common in electrodynamic problems and the way to incorporate them in the BCs across the interface is described in any standard textbook, see e.g., \cite{jackson}. On the other hand, the normal component ${P}_{S,\perp}^{(2)}$ corresponds to a perpendicular surface zero-thickness current, ${J}_{S,\perp}^{(2)}$, or equivalently, to a dipole layer pointing normal to the interface. Such a component is unusual and requires a more careful treatment. In the electrostatic limit, Jackson~\cite{jackson} derived the scalar electric potential for a {\em uniform} surface dipole layer oriented along the normal direction {\em in free space}~\cite{ftnt1}. Heinz~\cite{hnz} generalized this BC to an electrodynamic setting (i.e., for Maxwell´s Equations rather than Laplace´s Equation). 
Specifically, he modeled the surface sources in the zero thickness limit of a thin ``pill box'' in which the dipole layer is placed. 
Heinz's theory generalizes Jackson's BCs~\cite{jackson} in the following aspects:
\begin{enumerate}
\item The permittivity of the pill box $\varepsilon^{\prime}$ is allowed to be different from vacuum.
\item The surface dipole layer is allowed to be inhomogeneous (along the transverse coordinates).
\end{enumerate}

The BCs across the interface then read
\begin{eqnarray}
\Delta D_{\perp} &=& - \nabla_{\parallel} \cdot \mathbf{P}^{(2)}_S \equiv \varsigma^{(2)}, \label{eq:D_n} \\
\Delta \mathbf{E}_{\parallel} &=& - \frac{1}{\varepsilon^{\prime}} \nabla_\parallel {P}^{(2)}_{S,\perp} , \label{eq:E_t} \\
\Delta B_{\perp} &=& 0, \label{eq:B_n} \\
\Delta \mathbf{H}_{\parallel} &=& \mathbf{J}^{(2)}_S \times \hat{r}_{\perp} \equiv - 2 i \omega \textbf{P}^{(2)}_{S} \times \hat{r}_{\perp}, \label{eq:H_t}
\end{eqnarray}
where $\Delta$ denotes the difference of the variable across the interface. $\nabla_{\parallel}$ and $\hat{r}_{\perp}$ correspond to the tangential component of the vector derivative operator and the unit vector normal to the interface, respectively. We see that Eq.~(\ref{eq:D_n}), and Eqs.~(\ref{eq:B_n})-(\ref{eq:H_t}) are just the usual BCs found in standard textbooks (see e.g.,~\cite{jackson}). However, Eq.~(\ref{eq:E_t}) is more general. In particular, it shows that the spatial variation of the nonlinear polarization along the transverse coordinates results in discontinuity of $\mathbf{E}_{\parallel}$. Importantly, Eq.~(\ref{eq:E_t}) reduces to the standard BC when the normal component of $\mathbf{P}^{(2)}_{S}$ has zero gradient along the tangential direction, i.e., for $ \nabla_\parallel  {P}^{(2)}_{S,\perp} =0$. Moreover, note that the normal component of the electric field $E_{\perp}$  in Eq.~(\ref{eq:D_n}) becomes singular at the due to the presence of the dipole layer. Finally, Equation~(\ref{eq:E_t}) was also derived based on Green's function analysis in the context of planar structures by Sipe \cite{sipe86}. 

We now discuss the implementation of the BC given by Eq.~(\ref{eq:E_t}). To capture the unusual aspects of this general BC for SH radiation in a simple configuration, we choose to work with a $ \mathbf{P}^{(2)}_{S}$ given by
\begin{eqnarray}
\mathbf{P}^{(2)}_{S}=\varepsilon_0 \chi_{S,\perp,\perp,\perp}^{(2)}  {E}_{\perp}^{\omega}({\mathbf{r}}_{\parallel}) {E}_{\perp}^{\omega}( {\mathbf{r}}_{\parallel}) {\hat{{r}}}_{\perp},\label{p2_spc}
\end{eqnarray}
where $E_{\perp}^{\omega}( {\mathbf{r}}_{\parallel})$ is the normal component of the electric field at the FF. Such a polarization corresponds to a purely surface source, which was shown to be the dominant contribution to the $\bar{\bar{\chi}}^{(2)}_{S}$ tensor~\cite{wang09,jnl}. 

Note that $\chi^{(2)}_{S, \perp,\perp,\perp}$ is in general frequency dependent~\cite{rudnik,jnl}. However, the SH electric fields in all our results are normalized to their maximum values, thus, the actual value of $\chi^{(2)}_{S, \perp,\perp,\perp}$ is not important. Recall that the RHS of Eq.~(\ref{eq:E_t}) is non-zero only when $\mathbf{P}^{(2)}_{S}$ has non-zero gradient along the tangential direction, thus yielding a discontinuous parallel SH field, $\mathbf{E}^{2\omega}_{\parallel}$. This is the case in Eq.~(\ref{p2_spc}) when applied to any curved interface and a flat interface under oblique FF incidence.

The generalized BC in Eq.~(\ref{eq:E_t}) leads to an ambiguity since $\mathbf{P}^{(2)}_{S}$ in Eq.~(\ref{p2_spc}) depends on ${E}_{\perp}^{\omega}$  at the boundaries, which is inherently discontinuous due to the induced charges at the FF. Therefore, it is \textit{a-priori} not clear if ${E}_{\perp}^{\omega}$ should be evaluated on the metal or on the dielectric side of the interface. In order to resolve this seeming arbitrariness, Sipe \textit{et al.} argued that the FF fields in the nonlinear polarization are to be chosen {\em inside the metal}~\cite{sipe80,footnt1}. It was also shown that the resulting  dipole originating from $\mathbf{P}^{(2)}_{S}$ radiates as if placed outside the metal surface. The placement of this dipole layer outside the metal corresponds to $\varepsilon^{\prime}=\varepsilon_{bg}^{2\omega}$ in the generalized BC in Eq.~(\ref{eq:E_t}). Mizrahi \textit{et al.} established a proper correspondence between Heinz's and Sipe's derivation for planar slab geometries \cite{miz88}.

One of the main goals of this paper is to provide a simple method to calculate SHG without invoking Eq.~(\ref{eq:E_t}) and also to confirm its correspondence with the emergence of a SH dipole layer source outside the metal from a purely mathematical point of view for arbitrary geometries. We use a spectral decomposition method to achieve this.

\section{Spectral Decomposition method (SDM)}\label{sec:sec3}
We now outline our approach to calculate  the SHG based on the spectral decomposition method~\citep{DJB}. In short, SDM is an eigenmode expansion solution for the electrodynamic fields produced by an arbitrary current source even in the presence of a lossy resonator in an open system. In what follows, we briefly discuss the full electrodynamic formulation of SDM for non-magnetic and isotropic media. We then outline the full electrodynamic SDM solution for SHG arising from $\mathbf{P}^{(2)}_S$ [Eq.~(\ref{p2_spc})].

The SDM~\citep{DJB} is based on the electrodynamic analogue of the Lippmann-Schwinger equation. 
It begins with the standard inhomogeneous Maxwell's equations for a monochromatic wave, i.e., the vector Helmholtz equation, but rearranged to treat the response of the structure as another source term, namely,
\begin{equation} \label{sdm1}
\nabla\times(\nabla\times\mathbf{E}) - k_0^2 \varepsilon_{bg}\mathbf{E} = k_0^2 \left[\varepsilon(\mathbf{r}) - \varepsilon_{bg}\right]\mathbf{E} + i\omega\mu_0\mathbf{J}_f, 
\end{equation}
where $\varepsilon(\mathbf{r})$ is the spatial distribution of the permittivity and $k_0=\omega/c$. Here, we focus our attention on structures that have only two constituent materials, loosely defined as the inclusion ($\varepsilon_{in}$) and background ($\varepsilon_{bg}$). Thus, the free current source ($\mathbf{J}_f$)  and the bound (displacement) current response (proportional to [$\varepsilon_{in} - \varepsilon_{bg}$]) terms are treated on an equal footing. The Green's function solution to Eq.~(\ref{sdm1}) is the Lippmann-Schwinger equation, and has the form
\begin{equation} \label{sdm2}
\mathbf{E}(\mathbf{r}) = \mathbf{E}_0(\mathbf{r}) + k_0^2 \left[\varepsilon_{in}-\varepsilon_{bg} \right] \int \bar{\bar{\mathcal{G}}}_0 (|\mathbf{r} - \mathbf{r}^{\prime}|) \Theta(\mathbf{r}') \mathbf{E}(\mathbf{r}^{\prime})~ d\mathbf{r}^{\prime},
\end{equation}
where $\Theta(\bf{r}')$ is a Heaviside step function which is $1$ within the structure and $0$ elsewhere. The first term on the RHS of Eq.~(\ref{sdm2}) represents the electric field due to the free current source in a homogeneous background,
\begin{eqnarray} \label{e0sdm}
\mathbf{E}_0(\textbf{r})= i\omega\mu_0\int \bar{\bar{\mathcal{G}}}_0 (|\mathbf{r} - \mathbf{r}^{\prime}|) ~\mathbf{J}_f(\textbf{r}^{\prime})  d\mathbf{r}^{\prime}. 
\end{eqnarray}
The second term in Eq.~(\ref{sdm2}) captures the response of the system, and describes the field scattered by the structure. Due to the simple LHS of Eq.~(\ref{sdm1}), both terms in Eq.~(\ref{sdm2}) share the same simple dyadic Green's function for a homogeneous medium,
\begin{equation} \label{sdm3}
\bar{\bar{\mathcal{G}}}_0(|\mathbf{r} - \mathbf{r}^{\prime}|) = \left[\bar{\bar{\mathbf{I}}} - \frac{1}{k_0^2\varepsilon_{bg}} \nabla \nabla \cdot\right]  \mathcal{G}_0.
\end{equation}
$\mathcal{G}_0$ is available analytically, being the scalar Green's function for a homogeneous medium of premittivity $\varepsilon_{bg}$, which depends only on the dimension of the geometry investigated. Thus, $\mathbf{E}_0(\mathbf{r})$ in Eq.~(\ref{sdm2}) is relatively straightforward to find.

The efficient solution of the second term of the Lippmann-Schwinger Eq.~(\ref{sdm2}) is the key purpose of SDM. In quantum mechanics, the Lippmann-Schwinger equation is typically solved via either the Born series or simply the Born approximation. In electrodynamics, Eq.~(\ref{sdm2}) is the basis of the volume integral method of moments \cite{harrington1996field}, and the discrete dipole approximation \cite{purcell1973scattering,lakhtakia1990macroscopic}. In contrast, SDM uses Eq.~(\ref{sdm2}) as the basis of an {\em analytic} formulation, expanding the response of the structure by its eigenmodes. The eigenmodes are defined by the source-free version of the Lippmann-Schwinger equation Eq.~(\ref{sdm2}), 
\begin{equation} \label{sdm4}
s_l \mathbf{E}_l(\mathbf{r}) =k_0^2 \int \bar{\bar{\mathcal{G}}}_0 (|\mathbf{r} - \mathbf{r}'|) \Theta(\mathbf{r}') \mathbf{E}_l(\mathbf{r}')\, d\mathbf{r}',
\end{equation}
where the eigenvalue is 
\begin{eqnarray}\label{sdef}
s_l \equiv \frac{1}{\varepsilon_l - \varepsilon_{bg}}.
\end{eqnarray}
Here, we choose to fix the frequency ($k_0$) and vary the inclusion permittivity until Eq.~(\ref{sdm4}) is satisfied. Such eigenmodes are not the usual eigenmodes found in the literature, where the permittivity is fixed and frequency is taken as the eigenvalue, and instead correspond to a fixed frequency with the permittivity of the structure being the eigenvalue. The eigenvalue equation is thus defined by replacing $\epsilon_{in}$ of Eq.~(\ref{sdm2}) by $\varepsilon_l$ in Eq.~(\ref{sdef}), the latter corresponding to the eigen-permittivity of the mode. Note that $\varepsilon_l$ differs from, and is independent of the actual permittivity of the structure/inclusion ($\varepsilon_{in}$), and the set of all modes each with a different $\varepsilon_l$ provides a basis applicable to all possible $\epsilon_{in}$. For complete details related to the evaluation of the eigenvalues and normalized eigenmodes, see Ref.~\cite{2017arXiv170501747C}.

Once the eigenmode basis has been obtained, Eq.~(\ref{sdm2}) can be solved by projecting onto the set of eigenmodes of Eq.~(\ref{sdm4}). After some manipulation~~\cite{2017arXiv170501747C}, this procedure yields
\begin{equation} \label{sdm6}
|\mathbf{E}\rangle = |\mathbf{E}_0\rangle + \sum_{l\in\mathbb{N}} |\mathbf{E}_l\rangle \left(\frac{\varepsilon_{in} - \varepsilon_{bg}}{\varepsilon_l-\varepsilon_{in}} \right) \langle \mathbf{E}_l|\hat{\theta}| \mathbf{E}_0 \rangle,
\end{equation}
where for convenience, we have introduced the Dirac notation of quantum mechanics. Equation~(\ref{sdm6}) expresses the response of the system in terms of the eigenmodes, governed by two weights. First is the detuning of the inclusion's actual permittivity, $\varepsilon_{in}$, from the modal eigen-permittivity, $\varepsilon_{l}$. For $\varepsilon_l=\varepsilon_{in}$, one encounters a resonance of the structure~\cite{ftnt2}. Close to this resonance, the response of the system (for well-separated resonances) is governed by the mode $|\mathbf{E}_l\rangle$,  and the contribution from all the other modes and $|\mathbf{E}_0\rangle$ can be neglected. Second, $\langle\mathbf{E}_l|\hat{\theta}|\mathbf{E}_0\rangle$ determines the efficiency of excitation of each mode $|\mathbf{E}_l\rangle$ as a function of overlap of the specific shape of the illumination field with the mode $l$. Finally, Eq.~(\ref{sdm6}) demonstrates an important advantage of SDM, since the $|\mathbf{E}\rangle$ produced by the current source can be obtained without further simulation once the eigenmodes are known.


The quasi-static limit of SDM was used in the past to demonstrate the nano-focusing properties of surface plasmons~\cite{sdmap1,sdmap2,sdmap3,sdmap4,sdmap5} and to calculate SHG from metal nano-particles~\cite{sdmshg1,sdmshg2}. Here, we proceed a step further by applying SDM to describe surface plasmon assisted SHG in the electrodynamics limit within the so-called the undepleted pump approximation~\cite{boyd}. In our SHG calculations, the following  variables $\{\mathbf{E}(\mathbf{r})$, $|\mathbf{E}_0\rangle $, $|\mathbf{E}_{l}\rangle$, $\varepsilon_{(in, bg, l)}$, $\omega,~k_0\}$ are replaced by $\{\mathbf{E}^{2\omega}(\mathbf{r})$, $|\mathbf{E}_0^{2\omega}\rangle$, $|\mathbf{E}_{l}^{2\omega}\rangle$, $\varepsilon_{(in, bg, l)}^{2\omega}$, $2\omega,~2k_0\}$. The free current ($\mathbf{J}_{f}$) dependence of $|\mathbf{E}_0^{2\omega}\rangle$ in Eq.~(\ref{e0sdm}) is now given by $\mathbf{J}^{(2)} = -2 i\omega \mathbf{P}^{(2)}$. It is then straightforward to evaluate $|\mathbf{E}_0^{2\omega}\rangle$ from Eq.~(\ref{e0sdm}) using the free space Green's function. 
Our main observation is that, since the eigenmodes are obtained for a source-free  configuration, the tangential components of the eigenmodes, i.e., $\langle\mathbf{E}_{l,\parallel}^{2\omega}|$, are continuous across the interface. Thus, it follows from Eq.~(\ref{sdm6}) that the discontinuity of $\mathbf{E}_{\parallel}^{2\omega}$ across the interface for perpendicular surface current sources can arise \textit{only} from $|\mathbf{E}_0^{2\omega}\rangle$. In that sense, in SDM, the parameter $\varepsilon'$ in Eq.~(\ref{eq:E_t}) is naturally and unambiguously set to the background dielectric function ($\varepsilon^{2\omega}_{bg}$) by construction (without invoking the derivation of Sipe \textit{et al.} \cite{sipe80}).

In the following Section, we demonstrate the power of SDM  by considering SHG from  the simple geometry of a single metal cylinder.
\section{SHG from Single cylinder with surface nonlinearity: SDM implementation}\label{sec:sec4} 

We study SHG arising due to surface nonlinearity from an infinitely long metallic cylindrical inclusion (of permittivity 
$\varepsilon_{in} = \varepsilon_{cyl}$) of radius $a$ placed in a dielectric background. In this geometry,  $(\mathbf{r}_{\parallel},{r}_{\perp})$ take the form $(\theta, \rho)$ corresponding to cylindrical coordinates in two dimensions. We assume a normally incident ($\partial_z = 0$) in-plane TE polarized plane wave at FF. The total FF fields generated by the cylinder serve as the source for the SHG. The surface nonlinear polarization $\mathbf{P}^{(2)}_S$ is now placed at the interface $\rho=a$. We use vector cylindrical harmonics to project $\mathbf{J}^{(2)}$ on onto various angular orders $m$, giving
\begin{eqnarray} \nonumber
\mathbf{J}_{m}^{(2)}=-\frac{i\omega\varepsilon_0}{\pi}  \chi^{(2)}_{S,\perp\perp\perp}\left[ \int_0 ^{2\pi} \left[ E_{\rho}^{\omega}
\left(a,\theta^{\prime}\right) \right]^2 e^{-im\theta^{\prime}}  d\theta^{\prime} \right]  \\
\times e^{im\theta}~ \delta(\rho-a)~\hat{\rho}. \label{jm}
\end{eqnarray}
In writing Eq.~(\ref{jm}), we chose the FF fields from the metal, in accordance with Ref.~\cite{sipe80}. Once $\mathbf{J}_{m}^{(2)}$ is evaluated, it is then straightforward to calculate $|\mathbf{E}_0^{2\omega}\rangle$ from Eq.~(\ref{e0sdm}). The free space scalar Green's function in two-dimensions is given by the Hankel function of first kind $\mathcal{H}^{(1)}_0(k_{bg}^{2\omega}|\rho-\rho^{\prime}|)$ with $k_{bg}^{2\omega}=2\sqrt{\varepsilon_{bg}^{2\omega}}k_0$. We use Graf's addition theorem to shift $\mathcal{H}^{(1)}_0$ to the coordinate origin \cite{arfken}. The analytical solution of $|\mathbf{E}_0^{2\omega}\rangle$ is thus 
\begin{eqnarray} \nonumber
|\mathbf{E}_0^{2\omega}(\rho<a,\theta)\rangle &=&\sum_{m\in\mathbb{Z}}A_me^{im\theta} \left[ \frac{2m^2}{a\rho} \mathcal{J}_m(\rho k_{bg}^{2\omega}) \mathcal{H}_{m}^{(1)}(a k_{bg}^{2\omega})\hat{\rho} \right. \\ 
&+& \left. \frac{2im}{a} k_{bg}^{2\omega} \mathcal{J}_m^{\prime}(\rho k_{bg}^{2\omega})\mathcal{H}_{m}^{(1)}(a k_{bg}^{2\omega})\hat{\theta}\right], \label{e01} \\  \nonumber 
|\mathbf{E}_0^{2\omega}(\rho > a,\theta)\rangle&=&\sum_{m\in\mathbb{Z}}A_me^{im\theta} \left[ \frac{2m^2}{a\rho} \mathcal{J}_m(ak_{bg}^{2\omega}) \mathcal{H}_{m}^{(1)}(\rho k_{bg}^{2\omega})\hat{\rho} \right. \\ \label{e0out}
&+& \left. \frac{2im}{a} k_{bg}^{2\omega} \mathcal{J}_m(a k_{bg}^{2\omega})\mathcal{H}_{m}^{\prime(1)}(\rho k_{bg}^{2\omega})\hat{\theta}\right], 
\end{eqnarray} 
where $\mathcal{J}_m$ ($\mathcal{H}^{(1)}_m$) and $\mathcal{J}_m^{\prime}$ ($\mathcal{H}^{\prime(1)}_m$) correspond to the Bessel (Hankel) function of order $m$ and its derivative, respectively. The constant $A_m$ is given by
\begin{equation*}
A_m=\frac{ia}{8\varepsilon_{bg}^{2\omega}}\chi^{(2)}_{S,\perp,\perp,\perp} \left[ \int_0 ^{2\pi} \left[ E_{\rho}^{\omega}
\left(a,\theta^{\prime}\right) \right]^2 e^{-im \theta^{\prime}}  d\theta^{\prime} \right].
\end{equation*}

Having evaluated $|\mathbf{E}_0^{2\omega}\rangle$, the eigenmodes and eigen-permittivities need to be found in order to complete the solution in Eq.~(\ref{sdm6}). These are defined by Eq.~(\ref{sdm4}), which corresponds to a cylinder at resonance. For cylindrical geometries however, it is not necessary to solve Eq.~(\ref{sdm4}) directly, since solutions of the standard step-index fiber dispersion relation also produce eigen-permittivities that satisfy Eq.~(\ref{sdm4}). This transcendental equation requires a root search in the complex plane, but we employed an efficient method based on contour integration, guaranteed to capture all the roots \cite{chen2017robust}. The corresponding modal fields are given in Ref.~\cite{2017arXiv170501747C}, which also describes the necessary normalization and projection steps to use Eq.~(\ref{sdm6}), all performed analytically.
\begin{center}
\begin{figure}[h!]
\centering
\includegraphics[scale=0.175]{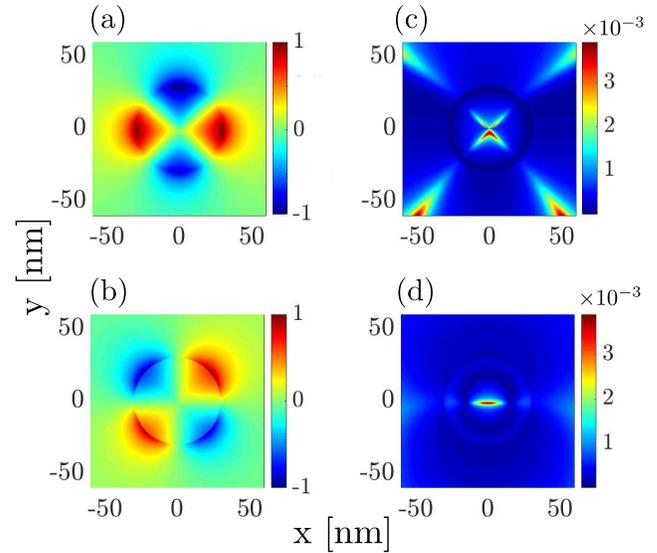}
\caption{(Color online) Normalized spatial distributions of SH electric fields for $\lambda_{FF}=700~$nm from a cylinder of  $a = 30$ nm, $\varepsilon_{bg} = 1$ and $\varepsilon_{cyl}$ is obtained from he Drude model with $\varepsilon_{\infty}=5$, $\omega_p=9.2$eV and $\gamma=0.021$eV, accounting for the angular orders $-4\leq m\leq 4$. (a) $D^{2\omega}_{\rho}$ and (b) $E^{2\omega}_\theta$ as calculated by nonlinear Mie-type solution for incident wavelength $\lambda_{FF} = 700$ nm. The relative error between the nonlinear Mie-type and SDM (with only the first radial mode, i.e, $l = 1$) solutions, namely, (c) $|D^{2\omega}_{\rho,Mie} - D^{2\omega}_{\rho,SDM}|/|D^{2\omega}_{\rho,Mie}|$ and (d) $|E^{2\omega}_{\theta,Mie} - E^{2\omega}_{\theta,SDM}|/|E^{2\omega}_{\theta,Mie}|$.}
\label{f1}
\end{figure}
\end{center}
\begin{center}
\begin{figure*}[t!] 
\centering
\includegraphics[scale=0.28]{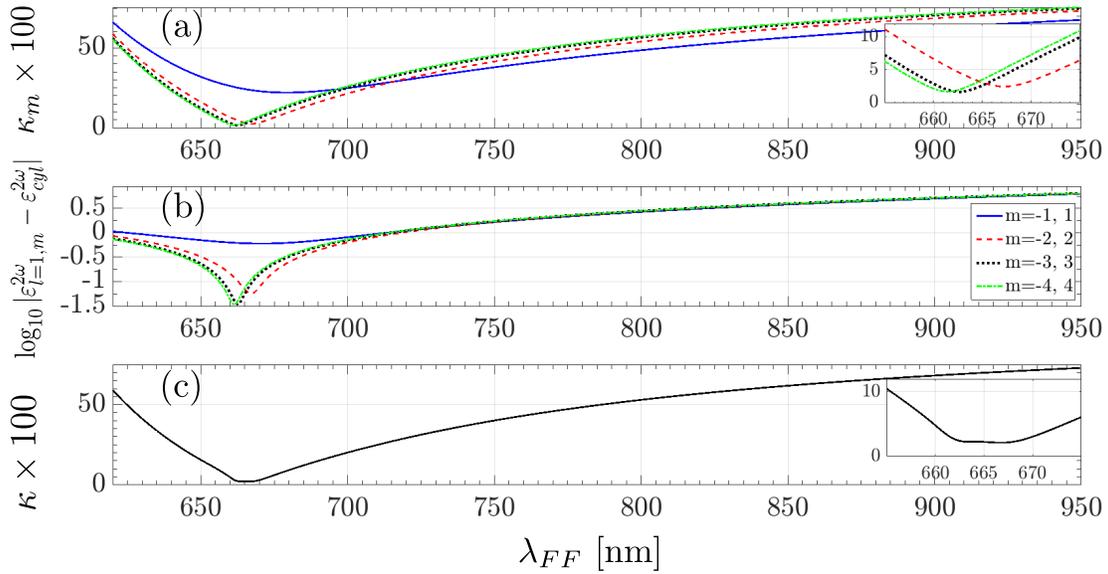}
\caption{(Color online) (a) $\kappa_m$ in percentages [as defined in Eq.~(\ref{eq_km})] (b) $\log_{10}|\varepsilon_{l=1}-\varepsilon_{cyl}|$ for various angular orders $m$. (c) $\kappa$ in percentages [as defined in Eq.~(\ref{eq_k})] as a function of $\lambda_{FF}$ for silver inclusion with $a=30$~nm. The inset in Fig.~\ref{f2}(a) [Fig.~\ref{f2}(c)] shows the behavior of $\kappa_m$ [$\kappa$] close to the resonance.  Other parameters are same as in Fig.~\ref{f1}.}
\label{f2}
\end{figure*}
\end{center}
\begin{center}
\begin{figure*}[t!]
\centering
\includegraphics[scale=0.28]{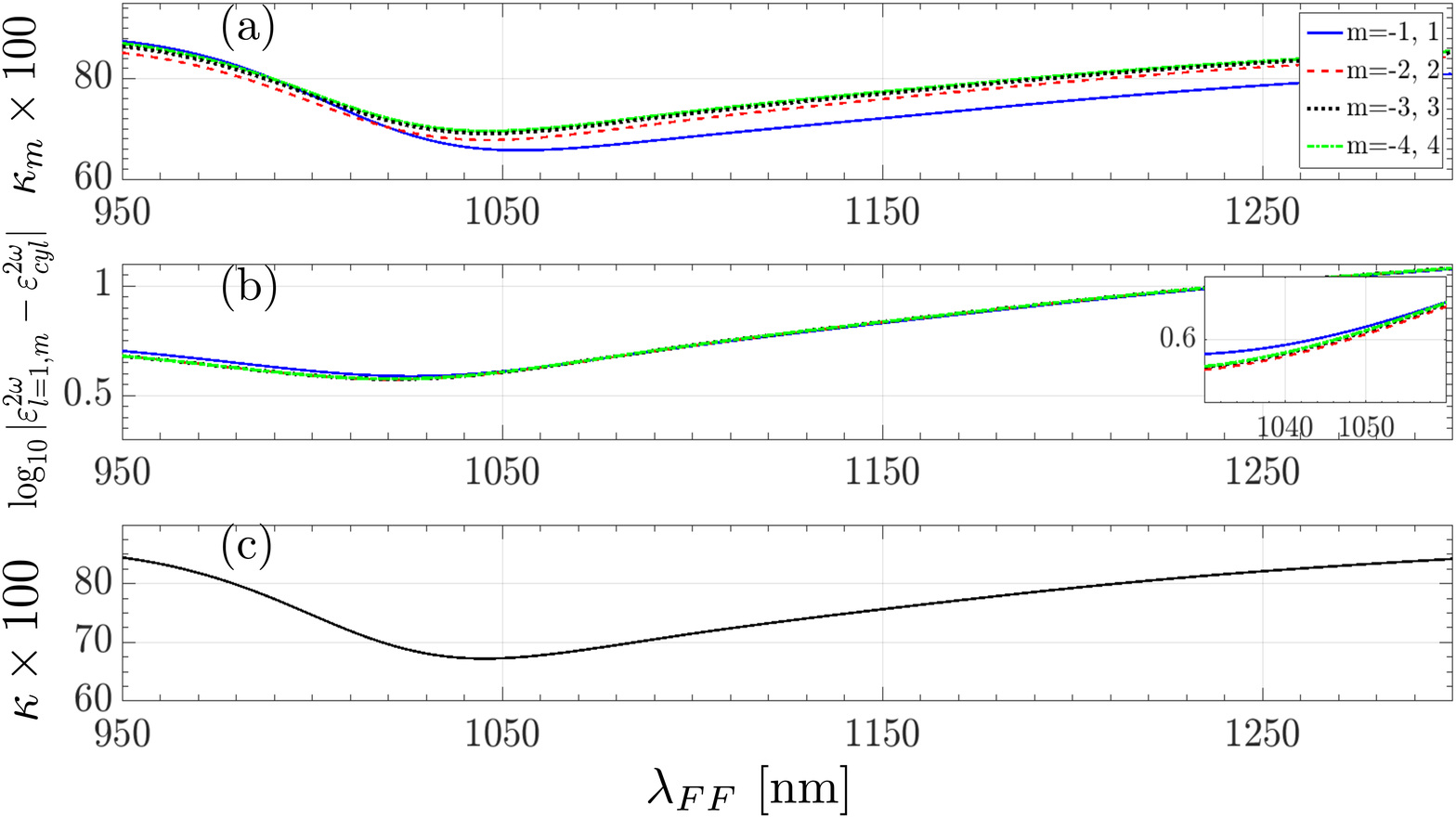}
\caption{(Color online) Same as in Fig.~\ref{f2} for a gold cylinder.} 
\label{f3}
\end{figure*}
\end{center}
In what follows, we compare the SH electric field solutions as evaluated by SDM to the nonlinear Mie-type solution (invoking Eq.~(\ref{eq:E_t}))~\cite{butet_coreshell}. We consider a typical cylindrical silver inclusion of radius $a=30$~nm, whose permittivity is given by the Drude model and the background medium is considered to be air. Other parameters are provided in the caption of Fig.~\ref{f1}. We consider the dominant contribution arising from the angular orders  $-4\leq m\leq4$ for the incident wavelength $\lambda_{FF}=700$~nm. We plot the normalized $D_{\rho}^{2\omega} $ and $E_{\theta}^{2\omega}$ as evaluated by nonlinear Mie-type solutions in Fig.~\ref{f1}(a) and Fig.~\ref{f1}(b), respectively.  We see that $D_{\rho}^{2\omega}$ is continuous, whereas $E_{\theta}^{2\omega}$ is discontinuous across the interface, as expected. The angular variation of $D_{\rho}^{2\omega}$ and $E_{\theta}^{2\omega}$ is dominated by the quadrapolar mode, i.e., $m=\pm2$. The relative errors between nonlinear Mie-type and SDM solution for $D_{\rho}^{2\omega}$ and $E_{\theta}^{2\omega}$ are shown in Fig.~\ref{f1}(c) and Fig.~\ref{f1}(d), respectively, showing excellent agreement between both solutions. It is important to note that this excellent agreement is achieved by incorporating only the first radial order (i.e., $l = 1$~\cite{ftnt4}) in computing $|\mathbf{E}^{2\omega}\rangle$ in Eq.~(\ref{sdm6}) for the angular modes $-4\leq m \leq 4$. Given the frequency of operation, the response of a generic plasmonic sub wavelength structure is dominated by the combination $|\mathbf{E}_0^{2\omega}\rangle$ and the first radial mode, which is usually plamonic in nature~\cite{2017arXiv170501747C}. Therefore, the first radial order would suffice to achieve such excellent agreement.

It follows from Eqs.~(\ref{e01})-(\ref{e0out}) that the normal component of $|\mathbf{E}_0^{2\omega}\rangle$, ${E}_{0,\rho}^{2\omega}$, is continuous across the interface, whereas the tangential component, $E_{0,\theta}^{2\omega}$ is discontinuous. Specifically, the $\mathbf{E}_{\parallel}^{2\omega}$-discontinuity is given by~\cite{ftnt3}
\begin{eqnarray} \nonumber
\Delta {E}_{0,\theta}^{2\omega} = \Delta {E}_{\theta}^{2\omega}= \sum_{m\in\mathbb{Z}}\frac{-im}{2\pi a\varepsilon^{2\omega}_{bg}} \chi^{(2)}_{S,\perp,\perp,\perp}  e^{im\theta} \\
\times \left[ \int_0 ^{2\pi} \left[ E_{\rho}^{\omega}
\left(a,\theta^{\prime}\right) \right]^2 e^{-im\theta^{\prime}}  d\theta^{\prime} \right].
\label{disce0}
\end{eqnarray}
Since the tangential components of the eigenmodes are continuous (as the eigenmodes are obtained from a source-free configuration), the discontinuity  $\Delta {E}_{\theta}^{2\omega}$ arises only from the free space contribution $\Delta {E}_{0,\theta}^{2\omega}$. Importantly, even though we did not employ surface-source terms in the boundary conditions, $\Delta {E}_{0,\theta}^{2\omega}$ [Eq.~(\ref{disce0})] has exactly the same form as in Eq.~(\ref{eq:E_t}) with $\varepsilon^{\prime}=\varepsilon^{2\omega}_{bg}$.

\section{The spectral dependence of discontinuity in ${E}_{\parallel}^{2\omega}$} \label{sec:sec5}
We now study the spectral dependence of the discontinuity in the tangential component of electric field $\Delta {E}_{\theta}^{2\omega}$ and exploit SDM to provide physical insight into it. Specifically, for a given angular order $m$, it can be seen from Eq.~(\ref{sdm6}) that close to the resonance of the first radial order ($\varepsilon^{2\omega}_{l=1,m} \rightarrow \varepsilon^{2\omega}_{cyl}$), the first weight factor  attains a large value. Hence, the mode $|\mathbf{E}_{l=1,m}^{2\omega}\rangle$ contributes dominantly to $|\mathbf{E}^{2\omega}\rangle$, while the contributions from $|\mathbf{E}_0^{2\omega}\rangle$ and other radial modes are negligible.  Thus, close to this resonance, the discontinuity $\Delta {E}_{\theta}^{2\omega}$ will be minimal. On the other hand, away from this resonance, the $|\mathbf{E}_0^{2\omega}\rangle$ contribution is significant, hence  $\Delta {E}_{\theta}^{2\omega}$ will be significant. Such a simple physical insight is not provided by the nonlinear Mie-type solution~\cite{butet_coreshell} and this is crucial in understanding the importance of the errors associated with the incorrect implementation of the generalized BC~(\ref{eq:E_t}). In order to quantify this trend, we define $\kappa_m$ 
as the ratio of the difference and the sum of $E_{m,\theta}^{2\omega}$ across the interface, namely,  
\begin{eqnarray} \label{eq_km}
\kappa_m &=&\left| \frac{ E_{m,\theta}^{2\omega}(\rho = a+\delta a,\theta)- E_{m,\theta}^{2\omega}(\rho = a-\delta a,\theta)}{ E_{m,\theta}^{2\omega}(\rho = a+\delta a,\theta)+E_{m,\theta}^{2\omega}(\rho = a-\delta a,\theta) } \right|,
\end{eqnarray}
where $\rho = a \pm \delta a$ correspond to a point outside/inside the cylinder with $\delta a/a \ll 1$. Note that unlike $\Delta {E}_{\theta}^{2\omega}$, which depends on $\theta$, $\kappa_m$ does not depend on $\theta$ as it is defined for a specific angular order. 

Figure~\ref{f2}(a) plots $\kappa_m$ as function of incident wavelength $\lambda_{FF}$ for a typical silver cylinder with $a=30~$nm  for $-4\leq m \leq 4$. In order to gain additional physical insight, we plot the first weight of Eq.~(\ref{sdm6}), i.e., $|\varepsilon_{l=1,m}^{2\omega}-\varepsilon_{cyl}^{2\omega}|$ in Fig.~\ref{f2}(b). The location of the minimum value attained by $|\varepsilon_{l = 1,m}^{2\omega}-\varepsilon_{cyl}^{2\omega}|$ defines the resonance of that particular angular order $m$. For example, the quadrapolar mode ($m=\pm2$) is resonant around $\lambda_{FF}=667$~nm and exactly at the same location $\kappa_{m=\pm2}$ attain a minimum value of about 2.5\%. $|\varepsilon_{l = 1,m}^{2\omega}-\varepsilon_{cyl}^{2\omega}|$ also determines the quality factor of the resonance, i.e., the smaller the minimum value of $|\varepsilon_{l = 1,m}^{2\omega}-\varepsilon_{cyl}^{2\omega}|$, the sharper the resonance. Importantly, away from the resonance, $\kappa_{m}$ takes large values ($>$50\%), meaning that the discontinuity becomes comparable to the field itself.

Since different angular orders $m$ have different weights in computing $E_{\theta}^{2\omega}$, it is not possible to determine the relative magnitude of $\Delta {E}_{\theta}^{2\omega}$ from $\kappa_m$ alone. Thus, we also define $\kappa$ as 
\begin{eqnarray}
\kappa = \text{max}\left| \frac{ E_{\theta}^{2\omega}(\rho = a+\delta a,\theta)- E_{\theta}^{2\omega}(\rho = a-\delta a,\theta)}{\text{max} \left[E_{\theta}^{2\omega}(\rho = a+\delta a,\theta)+E_{\theta}^{2\omega}(\rho = a-\delta a,\theta) \right]} \right|, \label{eq_k}
\end{eqnarray}
and plot its  spectral dependence in Fig.~\ref{f2}(c). We see that close to resonance, i.e., for $\lambda_{FF}=665~$nm, $\kappa$ attains a minimum value of about 2.5\%, but away from resonance it attains values higher than 50\%. For very large values of $\lambda_{FF}$, both $\kappa$ and $\kappa_m$ attain asymptotic values (not shown in the plot). 

Finally, we compare the magnitude of the discontinuity for different materials. 
We now consider a gold cylinder (see Fig.~\ref{f3}) and the permittivity of the gold  is obtained from the experimental data~\cite{J&C}. As before, we plot three quantities, namely, $\kappa_m$, $|\varepsilon_{l = 1,m}^{2\omega}-\varepsilon_{cyl}^{2\omega}|$ and $\kappa$ in Fig.~\ref{f3} and these quantities exhibit similar trends. However, the locations of the resonances and the minimum values attained by various quantities are substantially different, as expected. Indeed it is well known that gold is more lossy than silver. As a consequence, the minimum values attained by all the three quantities at the resonances for a gold inclusion are much higher than that of the silver inclusion. The minimum value attained by $\kappa$ for the gold inclusion is about 67\%, implying that the $\mathbf{E}^{2\omega}_{\parallel}$-discontinuity is significant throughout the spectrum.
\section{Numerical implementation}\label{sec:sec6}
Many studies have implemented the generalized BC in SH calculations using the nonlinear Mie-type solution~\cite{butet_coreshell}, the weak form of the differential equations~\cite{jnl}, surface integral methods~\cite{bem,miano,martan} and various other methods~\cite{mendez1,mendez2,panoiu,moreno}.

We now discuss the implementation of the generalized BC in the commercial numerical package COMSOL Multiphysics~\cite{comsol}. Having solved for FF fields using the scattering wave formulation, $J_{S,\perp}^{(2)}$ can be constructed following Eq.~(\ref{p2_spc}). However, the default surface current settings in the radio frequency (RF) module do not permit the surface current to have a normal component, so the effect of $J_{S,\perp}^{(2)}$(in general, any given $J_{S,\perp}$) is not accounted for~\cite{ftnt5}. Accordingly, modeling any surface nonlinearity (such as the models proposed in Refs.~\cite{ciraci1,ciraci2,scaloraPRA}) with a perpendicular component (i.e., including models more general than the one used here Eq.~(\ref{p2_spc})) based on COMSOL's {\em existing} settings is necessarily only partially correct. 

Instead of the default settings, one can choose to work with the weak form of the differential equations or redefine the boundary conditions. An alternative method to implement the generalized BC in Eq.~(\ref{eq:E_t}) is to approximate the surface with a very thin layer 
and use the external bulk current sources option. Note that the thin layer approximation is a numerical implementation of Heinz's derivation~\cite{hnz}. Following Sipe \textit{et al.}, the FF fields (needed for the calculation of $\mathbf{P}^{(2)}_S$) have to be chosen from the metal side of the interface and the thin layer (of thickness $t$) should have the permittivity of the background, i.e., $\varepsilon^{\prime}=\varepsilon_{bg}^{2\omega}$~\cite{sipe86}. Hence, this implementation results in the cylinder radius at SH being $a-t$. Thus, while this approach can be applied to most structures, it would fail to mimic the correct physics of complicated structures, such as kissing cylinders~\cite{alex}, as the structures at SH will always be non-touching, thus, modifying the response significantly~\cite{non-tchng}. Moreover, this implementation can be computationally expensive as the thin layer requires extremely fine numerical discretization.

In order to circumvent this limitation, and as a simpler alternative to adding the thin layer, we exploit the physical interpretation of $|\mathbf{E}_0^{2\omega}\rangle$ from the SDM. We use $|\mathbf{E}_0^{2\omega}\rangle$ [see Eqs.~(\ref{e0sdm})] generated by $J_{S,\perp}^{(2)}$ as our background {\em volume} incident field in the scattering field formulation instead of the surface source $J_{S,\perp}^{(2)}$ itself, see Appendix~\ref{app:Jperp_to_E0}. For the single cylinder case, we have used the analytical expressions of $|\textbf{E}_0^{2\omega}\rangle$ [see Eqs.~(\ref{e01})-(\ref{e0out})] in the scattered wave formulation. We have tested the numerical convergence and find the field distributions to be in excellent agreement with the nonlinear Mie-type solution (not shown). Furthermore, we note that when it is not possible to obtain an analytical expression for $|\mathbf{E}_0^{2\omega}\rangle$ for an arbitrary structure, one may rely on surface integral methods to evaluate $|\mathbf{E}_0^{2\omega}\rangle$ [see Eq.~(\ref{e0sdm})] arising from the surface current sources in free space/homogeneous background~\cite{BEMbook,bem}.

\section{Discussion and outlook} \label{sec:sec7}
In order to optimize the SHG efficiency from nano plasmonic structures, one needs to make sure that both FF and SH frequencies are tuned to different resonances supported by the structure. Our results show that if only the excitation (FF) is resonant, then, an incorrect implementation of the generalized BC~(\ref{eq:E_t}) can lead to severe errors in the computed SH near-fields. On the other hand, 
if the SH emission is resonant, then, the $\mathbf{E}_{\parallel}^{2\omega}$-discontinuity could be small (for instance, see Fig.~\ref{f2}) and thus, even if the $\mathbf{E}_{\parallel}^{2\omega}$-discontinuity is not implemented via Eq.~(\ref{eq:E_t}), the parallel component of electric field can still be reasonably accurate. However, the correct implementation of the generalized BC is crucial throughout the spectrum for gold inclusion. The effect of an incorrect implementation of the BC (e.g., a la COMSOL) on far-field quantities, such as the absorption and the scattering cross-sections, will depend sensitively on the type of nonlinearity tensor and geometry; such a study lies beyond the scope of the current paper.

Finally, it is clear that the effects demonstrated here for SHG from a single metal wire with a simplistic second-order polarization tensor apply also for other structures (e.g., 3D nanostructures, waveguides etc.~\cite{zayat_rev}), as well as for more sophisticated second-order optical effects, such as three wave mixing~\cite{tal1,tal2}. It is very easy to incorporate the models for SH sources in SDM which are more general than the one used in Eq.~(\ref{p2_spc}), i.e., including the other components of $\chi^{(2)}_{S}$, nonlocal bulk effects, etc. We hope that future studies will indeed adopt the formulation described here.
%
\appendix
\section{Sources} \label{apenA_sources}
We now obtain the relations between the surface and volume charge accumulations generated by the nonlinear surface polarization $\mathbf{P}^{(2)}_{S}$. The surface ($\sigma^{(2)}$) and volume ($\rho^{(2)}$) charge are related via $\int \sigma^{(2)} dS = \int \rho^{(2)} dV$. In order to relate them with a differential relation, we use the usual continuity relation $\partial_t \rho^{(2)} = -\nabla \cdot \mathbf{J}^{(2)}$ (with current source $\textbf{J}^{(2)}$ as the external source that arises from SH polarization,  $\mathbf{J}^{(2)}=-2i\omega\mathbf{P}^{(2)}$). $\sigma^{(2)}$ and $\rho^{(2)}$ in terms of $\mathbf{P}^{(2)}$ are given by
\begin{eqnarray} 
\rho^{(2)} = -\nabla \cdot \mathbf{P}^{(2)}, \quad \sigma^{(2)}= \mathbf{P}^{(2)} \cdot  {r}_{\perp}, \label{appenA_eq1}
\end{eqnarray}
respectively. Note that Eq.~(\ref{appenA_eq1}) is very general, however, it does not hold for the surface cases such as Eq.~(\ref{p2_gen}). We now derive the surface $\sigma^{(2)}$ and the volume $\rho^{(2)}$ charge accumulations arising from such surface sources. Integrating $\rho^{(2)}$ in Eq.~(\ref{appenA_eq1}) over the volume and invoking Eq.~(\ref{p2_gen}) gives us
\begin{eqnarray} \nonumber 
\int \rho^{(2)} dV  &=& -\int \nabla \cdot \mathbf{P}^{(2)} dV \\
&=& -\int \left[ \nabla_{\parallel} \cdot \textbf{P}^{(2)}_{S,\parallel}(\mathbf{r}_{\parallel})  \right] \delta( {r}_{\perp}) dV {\nonumber} \\ &-&\int  \textbf{P}^{(2)}_{S,\perp}(\mathbf{r}_{\parallel})d\mathbf{r}_{\perp} \int  \partial_{\perp}  \delta( {r}_{\perp}) d\mathbf{r}_{\perp} \label{appenA_eq2}\\
&=& -\int  \left[ \nabla_{\parallel} \cdot \mathbf{P}^{(2)}_{S}(\mathbf{r}_{\parallel}) \right] ~\delta( {r}_{\perp})~ dV \label{appenA_eq3} \\
=\int \sigma^{(2)} dS&=& \left. -\int \left[ \nabla_{\parallel} \cdot \mathbf{P}^{(2)}_{S}(\mathbf{r}_{\parallel}) \right|_{ {r}_{\perp}=0}\right] ~ dS.   \label{eq7} 
\end{eqnarray}
Note that in arriving at Eq.~(\ref{appenA_eq3}) from Eq.~(\ref{appenA_eq2}) we have employed the relation $ \int \partial_{\perp} \delta({r}_{\perp}) d {r}_{\perp}=0 $. Thus, necessarily, the surface and volume charges for the surface sources are defined by $\varrho^{(2)}$ and $\varsigma^{(2)}$, respectively, are given by 
\begin{eqnarray} \label{sigma-rho-diff}
\varrho^{(2)} = -\left[\nabla_{\parallel} \cdot \mathbf{P}^{(2)}_S \right] \delta( {r}_{\perp}),\quad\varsigma^{(2)}=-\nabla_{\parallel}\cdot \mathbf{P}^{(2)}_S.
\end{eqnarray}
It is important to note that Eq.~(\ref{appenA_eq1}) becomes Eq.~(\ref{sigma-rho-diff}) for the surface sources.

\section{Scattered wave formulation for a current source} \label{app:Jperp_to_E0}
Since the \textit{existing} COMSOL settings do not account for $J_{S,\perp}$, we would like to employ the insights provided by SDM to overcome this problem. Specifically, we now show how to derive $|\mathbf{E}_{0}^{2\omega}\rangle$ from $J_{S,\perp}$, and then use it in the scattered wave formulation. 
 
Helmholtz's equation with a free current source $J_{S,\perp}$ reads as 
\begin{equation}\label{apndx_b1}
\nabla\times(\nabla\times\mathbf{E}^{2\omega})-k^{2}(\mathbf{r})\mathbf{E}^{2\omega}=2i\mu_0\omega{J}_{S,\perp}^{(2)},\quad k^2(\mathbf{r})=4k^2_0\varepsilon^{2\omega}(\mathbf{r}).
\end{equation}
Now $\mathbf{E}^{2\omega}$ can be decomposed as
\begin{eqnarray}\label{apndx_b2}
\mathbf{E}^{2\omega}=\mathbf{E}_{bg}^{2\omega}+\mathbf{E}_{sc}^{2\omega},
\end{eqnarray}
where $\mathbf{E}_{bg}^{2\omega}$ and $\mathbf{E}_{sc}^{2\omega}$ correspond to the field originating from the current source ${J}_{S,\perp}^{(2)}$ in an homogeneous background and the field scattered from the structure, respectively. Thus, $\mathbf{E}_{bg}^{2\omega}$ obeys
\begin{eqnarray}\label{apndx_b3}
\nabla\times(\nabla\times\mathbf{E}_{bg}^{2\omega})-(k^{2\omega}_{bg})^2 \mathbf{E}_{bg}^{2\omega} = 2i\omega \mu_0 J_{S,\perp}^{(2)}.
\end{eqnarray}
Note that the above equation in its integral form is the same as Eq.~(\ref{e0sdm}) when written at SH, thus, $\mathbf{E}_{bg}^{2\omega}$ is identical to $|\mathbf{E}_{0}^{2\omega}\rangle$. Now, comparing Eq.~(\ref{sdm6}) to Eq.~(\ref{apndx_b2}) reveals that the scattered field $\mathbf{E}_{sc}^{2\omega}$ corresponds to the second term of RHS in Eq.~(\ref{sdm6}). Rewriting Eq.~(\ref{apndx_b1}) in terms of $\mathbf{E}_{sc}^{2\omega}$ by invoking Eqs.~(\ref{apndx_b2})-(\ref{apndx_b3}) and replacing $\mathbf{E}_{bg}^{2\omega}$ by $|\mathbf{E}_{0}^{2\omega}\rangle$ yields
\begin{eqnarray}\label{apndx_b4}
\nabla\times(\nabla\times\mathbf{E}_{sc}^{2\omega})-k^{2}(\mathbf{r})\textbf{E}_{sc}^{2\omega}= [k^2(\mathbf{r})-(k_{bg}^{2\omega})^2]\mathbf{E}_{0}^{2\omega}.
\end{eqnarray}
Eq.~(\ref{apndx_b4}) is nothing but what is known as the scattered wave formulation, so that now, $\mathbf{E}_{sc}^{2\omega}$ can be evaluated from $|\mathbf{E}_{0}^{2\omega}\rangle$ instead of $J_{S,\perp}^{(2)}$.

\section*{Funding Information}
KNR, PYC and YS were partially supported by Israel Science Foundation (ISF)
(899/16). YS acknowledges the financial support from the People Programme (Marie Curie Actions) of the European Union’s Seventh Framework Programme (FP7/2007-2013) under REA grant (333790) and the Israeli National Nanotechnology Initiative. AIFD acknowledges funding from EU Seventh Framework Programme under Grant Agreement FP7-PEOPLE-2013-CIG-630996, and the Spanish MINECO under contract FIS2015-64951-R.
\bibliography{ref}

\begin{thebibliography}{10}
\bibitem{ftnt1}
Ref.~\cite{jackson} demonstrates that the perpendicular component of the
  electric field becomes singular at the boundary due to the presence of the
  dipole layer source. Indeed, the electric potential is discontinuous in this
  case, so that its derivative (the electric field) diverges.
\bibitem{footnt1}
Note that apart from this convention, the FF fields can also be chosen outside,
  but then $\chi^{(2)}_{S,\perp,\perp,\perp}$ has to be multiplied by the
  appropriate factor of the dielectric functions \citep{wang09}.
\bibitem{ftnt2}
Note that for a realistic lossy/open system, $\varepsilon_l$ is complex with a
  imaginary component corresponding to gain. Thus, $\varepsilon_l$ can never
  equal the $\varepsilon_{in}$ of a passive medium, so the weight factor is
  never infinite.
\bibitem{ftnt3}
Note that the discontinuity decreases with $m$ because of the weaker excitation
  of the higher-order $m$ modes.
\bibitem{ftnt5}
COMSOL Multiphysics projects the surface current density onto the boundary
  surface and neglects its normal component.
\end{thebibliography}

\ifthenelse{\equal{\journalref}{ol}}{%
\clearpage
\bibliographyfullrefs{ref}
}{}

\end{document}